\def\dir{./}
\documentclass[useAMS,usenatbib]{\dir mn2e}
\usepackage[fleqn]{amsmath}
\usepackage{amssymb}
\usepackage[super]{nth}
\usepackage{natbib}
\usepackage{graphicx}
\usepackage{float}
\usepackage{mathrsfs}
\usepackage{mathtools}
\usepackage{multirow}
\usepackage[usenames,dvipsnames]{color}
\usepackage{\dir aas_macros}
\usepackage{comment}
\usepackage{subfigure}
\usepackage{hyperref}
\usepackage{mathrsfs}
\usepackage{epstopdf, dcolumn}
\DeclareGraphicsExtensions{.pdf}
\usepackage{wasysym}
\usepackage{dashrule}
\usepackage{xcolor}
\usepackage{arydshln}
\usepackage{threeparttable}
\usepackage[export]{adjustbox}
\usepackage[figuresleft]{rotating}

\newlength\replength
\newcommand\repfrac{.33}

\newcommand\rulewidth{.6pt}
\newcommand\tdashfill[1][\repfrac]{\cleaders\hbox to \replength{%
		
\smash{\rule[\arraystretch\ht\strutbox]{\repfrac\replength}{\rulewidth}}}\hfill}

\newcommand\tdotfill[1][\repfrac]{\cleaders\hbox to \replength{%
		
\smash{\raisebox{\arraystretch\dimexpr\ht\strutbox-.1ex\relax}{.}}}\hfill}

\newcommand{\appropto}{\mathrel{\vcentre{
			\offinterlineskip\halign{\hfil$##$\cr
				
\propto\cr\noalign{\kern2pt}\sim\cr\noalign{\kern-2pt}}}}}

\newcommand\lsim{\mathrel{\rlap{\lower4pt\hbox{\hskip1pt$\sim$}}
        \raise1pt\hbox{$<$}}}
\newcommand\gsim{\mathrel{\rlap{\lower4pt\hbox{\hskip1pt$\sim$}}
        \raise1pt\hbox{$>$}}}
\newcommand{\Rom}[1]{\uppercase\expandafter{\romannumeral #1}}
\newcommand{\rom}[1]{\lowercase\expandafter{\romannumeral #1}}
\newcommand{\htwo}{\mathrm{H}_2}
\newcommand{\hone}{\mathrm{H}\textsc{i}}

\newcommand{\tocm}{{$\mathrm{21cm}\textsc{fast}$}}
\newcommand{\tocmmc}{{$21\textsc{cmmc}$}}
\newcommand{\lya}{{Lyman-$\alpha$}}
\newcommand{\mvir}{M_{\rm h}}
\newcommand{\msol}{{\rm M}_\odot}
\newcommand{\rmd}{{\rm d}}

\defcitealias{Qin2020arXiv200304442Q}{Paper-I}

\begin{document}

\title[minihalos in 21-cm PS]{A tale of two sites - \Rom{2}: Inferring the 
properties of minihalo-hosted galaxies with upcoming 21-cm interferometers}
\author[Qin et al.]{Yuxiang Qin$^{1}$\thanks{E-mail: Yuxiang.L.Qin@gmail.com}, 
Andrei Mesinger$^1$, Bradley Greig$^{2,3}$ and Jaehong Park$^{1,4}$\\
	$^{1}$Scuola Normale Superiore, Piazza dei Cavalieri 7, I-56126 Pisa, 
Italy\\
	$^{2}$School of Physics, University of Melbourne, Parkville, VIC 3010, 
Australia\\
	$^{3}$ARC Centre of Excellence for All Sky Astrophysics in 3 Dimensions 
(ASTRO 3D)\\
    $^{4}$ School of Physics, Korea Institute for Advanced Study (KIAS), 85 Hoegiro, Dongdaemun-gu, Seoul 02455, Republic of Korea}
\maketitle
\label{firstpage}

\begin{abstract}
 The first generation of galaxies is expected to form in minihalos, accreting gas through ${\rm H}_2$ cooling, and possessing unique properties.  Although unlikely to be directly detected in UV/infrared surveys, the radiation from these molecular-cooling galaxies (MCGs) could leave an imprint in the 21-cm signal from the Cosmic Dawn.
        Here we quantify their detectability with upcoming radio interferometers.
We generate mock 21-cm power spectra using a model for both MCGs as well as more massive, atomic-cooling galaxies (AGCs), allowing both populations to have different properties and scaling relations.
The galaxy parameters are chosen so as to be consistent with: (i) high-redshift UV luminosity functions; (ii) the upper limit on the neutral fraction from QSO spectra; (iii) the Thomson scattering optical depth to the CMB; and (iv) the timing of the recent putative EDGES detection.
The latter implies a significant contribution of MCGs to the Cosmic Dawn, if confirmed to be cosmological.
We then perform Bayesian inference on two models including and ignoring MCG contributions. 
	Comparing their Bayesian evidences, we find a strong 
	preference for the model including MCGs, despite the fact that it has more free parameters. This suggests that if MCGs indeed play a significant role in the Cosmic Dawn, it should be possible to infer their properties from upcoming 21-cm power spectra.
        Our study illustrates how these observations can discriminate among uncertain galaxy formation models with varying complexities, by maximizing the Bayesian evidence.
\end{abstract}

\section{Introduction}

The first galaxies of our Universe are expected to form out of pristine gas, cooling inside so-called ``minihalos'' (with mass $\mvir\sim$ $10^6$ -- $10^8$ $M_\odot$) via rotational-vibrational transitions of H$_2$ (e.g. \citealt{Haiman1996ApJ...467..522H,Haiman1997ApJ...476..458H,Yoshida2003ApJ...598...73Y,Yoshida2006}).
The first episodes of star formation, evolution, and feedback inside these first-generation, molecular-cooling galaxies (MCGs) can be very different from later generations that were mostly built-up out of pre-enriched material inside deeper potential wells
(e.g. \citealt{Haiman1999,Tumlinson2000ApJ...528L..65T,Abel2002Sci...295...93A,Schaerer2002A&A...382...28S,Bromm2004ARA&A..42...79B,Yoshida2006,McKee2008ApJ...681..771M,Whalen2008ApJ...682...49W,Turk2009Sci...325..601T,Heger2010ApJ...724..341H,Wise2012ApJ...745...50W,Xu2016ApJ...833...84X,Kimm2016}).
Moreover, star formation inside MCGs is expected to be transient, 
tapering off as a growing Lyman-Werner (LW)
background starts to effectively photodissociate $\htwo$ (e.g.
\citealt{Johnson2007ApJ...665...85J,Ahn2009ApJ...695.1430A,Holzbauer2012MNRAS.419..718H,Fialkov2013MNRAS.432.2909F,Jaacks2018,Schauer2019MNRAS.484.3510S}).

Unfortunately, MCGs are likely too faint to observe directly using UV or infrared telescopes in the foreseeable future (e.g. \citealt{OShea2015ApJ...807L..12O,Xu2016ApJ...833...84X}).
Their transient nature also
makes low-redshift detection or even searching for stellar relics in the nearby Universe very challenging
\citep{Beers2005ARA&A..43..531B,Tornatore2007MNRAS.382..945T,Nagao2005ApJ...631L...5N,Nagao2008ApJ...680..100N,Lai2008ApJ...681.1524L,Suda2008PASJ...60.1159S,Roederer2014AJ....147..136R,Liu2020arXiv200615260L}.

A promising alternative is to study MCGs through the imprint their radiation fields leave in the
intergalactic medium (IGM; 
\citealt{Fialkov2013MNRAS.432.2909F,Mirocha2017,Munoz_2019}).
In the standard hierarchical structure formation paradigm, there should have existed a period at the start of the Cosmic Dawn in which the radiation backgrounds were dominated by MCGs.  If we can observe IGM properties at a high enough redshift, we could indirectly study the properties of MCGs (e.g.
\citealt{Ciardi2006MNRAS.366..689C,McQuinn2007MNRAS.377.1043M,Ahn2012ApJ...756L..16A,Visbal2015MNRAS.453.4456V,Miranda2017MNRAS.467.4050M,Mesinger2012MNRAS.422.1403M,Koh2018MNRAS.474.3817K}).

Luckily, the cosmic 21-cm signal is set to revolutionize our understanding of the early Universe (for a recent review, see \citealt{Mesinger10.1088/2514-3433/ab4a73}).  Sourced by the spin-flip transition of neutral hydrogen, the cosmic 21-cm signal is sensitive to the ionization and thermal state of the IGM.  These are in turn determined by the ionizing, soft UV and X-ray emission from the first galaxies.  Therefore, the high-redshift 21-cm signal should encode information about the birth, disappearance, spatial distribution, and typical spectral energy distributions (SEDs) of MCGs.

Many experiments are striving to measure the signal. These can be broadly divided into global signal experiments and interferometers measuring 21-cm fluctuations.  The former includes
the Shaped Antenna 
measurement of the background RAdio Spectrum (SARAS; 
\citealt{Singh2018ExA....45..269S}), the 
Large-aperture Experiment to Detect the Dark Age (LEDA; 
\citealt{Price2018MNRAS.478.4193P}), Probing Radio Intensity at high-Z from 
Marion (PRI$^{\rm Z}$M; \citealt{Philip2019JAI.....850004P}), and the 
Experiment to Detect the Global EoR (i.e. Epoch of Reionization) Signature 
(EDGES; \citealt{Bowman2018}). The latter has recently claimed a detection with 
an absorption feature at $z{\sim}17$, inciting much debate as to its 
cosmological origin (e.g. 
\citealt{Hills2018arXiv180501421H,Bradley2019ApJ...874..153B,Sims2019MNRAS.488.2904S,Mirocha2019MNRAS.483.1980M,Munoz2018Natur.557..684M,Fialkov2018PhRvL.121a1101F,Ewall-Wice2018ApJ...868...63E,Mebane2019arXiv191010171M,Qin2020arXiv200304442Q}). 
Existing interferometers, such as the Low-Frequency Array 
(LOFAR\footnote{\url{http://www.lofar.org/}}; 
\citealt{vanHaarlem2013A&A...556A...2V,Patil2017ApJ...838...65P}), the 
Murchison Widefield Array (MWA\footnote{\url{http://www.mwatelescope.org/}}; 
\citealt{Tingay2013PASA...30....7T,Beardsley2016ApJ...833..102B}) and the 
Precision Array for Probing Epoch of Reionisation 
(PAPER\footnote{\url{http://eor.berkeley.edu}}; \citealt{Parsons_2010}), are 
focusing on measuring the 21-cm power spectrum, generally at $z<10$.
 These instruments are serving as precursors and pathfinders for 
the next-generation radio telescopes: the Hydrogen Epoch of Reionization 
Arrays (HERA\footnote{\url{http://reionization.org/}}; 
\citealt{DeBoer2017PASP..129d5001D}) and the Square 
Kilometre Array (SKA\footnote{\url{https://www.skatelescope.org/}}), which 
promise to deliver 3-dimensional imaging and a high S/N measurement of the 21-cm power spectrum (PS) out to $z\lsim$ 20--30.

However, even with a clean detection of the signal, it is not obvious that we can claim to have detected MCGs.  Given how little we know about high-redshift galaxies, there could be many degeneracies in theoretical models.  {\it Would we be able to confidently extract the imprint of MCGs from the signal, and distinguish them from more evolved, second-generation galaxies?}
Bayesian inference provides us with a clean framework to answer such a question.  Specifically, Bayesian evidence allows us to perform model selection, quantifying if data prefers one theoretical model over another.  It has a built-in Occam's razor factor, penalizing additional model complexity unless explicitly required by the data (for a recent review of Bayesian inference in astronomy, see \citealt{Trotta2017arXiv170101467T}).

In this work, we quantify the detectability of MCGs from a mock measurement of the cosmic 21-cm PS, expected from a 1000h integration with SKA1-low.  Our mock signal is generated by self-consistently following the evolution of both MCGs and more massive atomically-cooled galaxies (ACGs), as described in \citet[][hereafter \citetalias{Qin2020arXiv200304442Q}]{Qin2020arXiv200304442Q}.  From this mock observation, we infer the properties of the underlying galaxies using a model having only a population of ACGs, and a model allowing for both populations: ACGs and MCGs.  We compute the Bayes factor of these two models, quantifying if the mock observation provides sufficient evidence for an additional population of MCGs.

This paper is organized as follows. We briefly summarize our model in Section 
\ref{sec:models}.  In Section \ref{sec:mock}, we present our mock observation, chosen so that the timing of the global signal is consistent with the putative EDGES detection.
In Section 
\ref{sec:constraint}, we perform parameter inference using two galaxy models, presenting the corresponding Bayesian evidence.
Finally, we 
conclude in Section \ref{sec:conclusion}. In this work, we adopt the following cosmological 
parameters: 
($\Omega_{\mathrm{m}}, \Omega_{\mathrm{b}}, \Omega_{\mathrm{\Lambda}}, h, \sigma_8, n_s $ = 0.31, 
0.048, 0.69, 0.68, 0.81, 0.97),  consistent with \textit{Planck} \citep{Planck2016A&A...594A..13P,Planck2018arXiv180706209P}).

\section{Characterizing galaxies at Cosmic Dawn}
\label{sec:models}

To model the 21-cm signal we use the public code {\tocm}\footnote{\url{https://github.com/21cmfast/21cmFAST}} \citep{Mesinger2007ApJ...669..663M, Mesinger2011MNRAS.411..955M} with the latest update from \citetalias{Qin2020arXiv200304442Q}. In \citetalias{Qin2020arXiv200304442Q},
we extended the galaxy models of {\tocm} to include a separate population of MCGs, with properties independent to those of ACGs.  Here we briefly summarize our procedure for characterizing these galaxies and their corresponding emissivities; for more details, please see \citetalias{Qin2020arXiv200304442Q}.

We define two distinct galaxy populations on the basis of the cooling channel through which they obtained the bulk of their gas  -- ACGs and MCGs.  These two populations are defined via exponential ``window functions'' over the halo mass ($\mvir$) function, ${\rmd n}/{\rmd \mvir}$ (for an in-depth discussion of this choice, see \citetalias{Qin2020arXiv200304442Q}).  Specifically, the number density of actively star-forming galaxies is
\begin{equation}\label{eq:hmf}
\phi = \frac{\rmd n}{\rmd \mvir} \times
\begin{cases}
\exp\left({-\dfrac{M_{\rm crit}^{\rm atom}}{\mvir}}\right)\\
\exp\left({-\dfrac{M_{\rm crit}^{\rm 
mol}}{\mvir}}\right)\exp\left({-\dfrac{\mvir}{M_{\rm crit}^{\rm cool}}}\right)
\end{cases},
\end{equation}
where the superscripts ``atom" and ``mol" are used to distinguish ACGs and 
MCGs, respectively, as they are allowed to have different properties and scaling relations.

We see from equation (\ref{eq:hmf}) that the occupancy fraction of ACGs starts dropping below a characteristic mass scale of
\begin{equation}\label{eq:m_crit_atom}
M_{\rm crit}^{\rm atom} = \max\left[ M_{\rm crit}^{\rm cool},  M_{\rm crit}^{\rm ion}, M_{\rm crit}^{\rm SN} \right] ~ .
\end{equation}
Here we account for three physical processes that can suppress star formation:
(i) inefficient cooling, $M_{\rm crit}^{\rm cool}$ (corresponding to a virial temperature of  ${\sim}10^4$K; \citealt{Barkana2000}); (ii) photoheating feedback from inhomogeneous reionization, $M_{\rm crit}^{\rm ion}$ (e.g. \citealt{Sobacchi2014MNRAS.440.1662S}; see also 
\citealt{Efstathiou1992MNRAS.256P..43E,Shapiro1994ApJ...427...25S,Thoul1996ApJ...465..608T,Hui1997MNRAS.292...27H}); and (iii) supernova feedback\footnote{Following \citetalias{Qin2020arXiv200304442Q}, here we also assume $M_{\rm crit}^{\rm SN}$ is smaller than the other relevant mass scales, so that we can maximize the importance of MCG and thus match the timing of the putative EDGES detection.  Note that SNe feedback could still be responsible for the power-law scaling of the stellar to halo mass relation if star formation is feedback-limited (e.g. \citealt{Wyithe2013MNRAS.428.2741W}).}, $M_{\rm crit}^{\rm SN}$ \citep{Haiman1999,Wise2007,DallaVecchia2008,DallaVecchia2012,Hopkins2014MNRAS.445..581H,Keller2014,Kimm2016,Hopkins2017}.

On the other hand, the occupancy fraction of MCGs picks up below the atomic cooling threshold, $M_{\rm crit}^{\rm cool}$, and extends down to
\begin{equation}\label{eq:m_crit_mol}
M_{\rm crit}^{\rm mol} = \max\left[ M_{\rm crit}^{\rm diss},  M_{\rm crit}^{\rm ion}, M_{\rm crit}^{\rm SN} \right] ~ ,
\end{equation}
where the additional term, $M_{\rm crit}^{\rm diss}$, accounts for the cooling efficiency of ${\htwo}$ in the presence of an inhomogeneous LW background (e.g. \citealt{Machacek2001ApJ...548..509M, Draine1996ApJ...468..269D,Johnson2007ApJ...665...85J,Ahn2009ApJ...695.1430A,Wolcott-Green2011MNRAS.418..838W,Holzbauer2012MNRAS.419..718H,Visbal2015MNRAS.453.4456V}).  

We adopt power-law relations for the stellar ($M_*$) to halo mass ratio 
\citep{Moster2013MNRAS.428.3121M,Sun2015,Mutch2016,Ma2018,Tacchella2018ApJ...868...92T,Behroozi2019MNRAS.488.3143B,Yung2019MNRAS.490.2855Y}
\begin{equation}\label{eq:m*}
\frac{M_*}{\mvir} =  \frac{\Omega_{\mathrm{b}}}{\Omega_{\mathrm{m}}}\times\min\left[1, \begin{cases}
f_{*,10}^{\rm atom}\left(\frac{\mvir}{10^{10}\msol}\right)^{\alpha_*}\\
f_{*,7}^{\rm mol}\left(\frac{\mvir}{10^{7}\msol}\right)^{\alpha_*}
\end{cases}
\right],
\end{equation}
where three free parameters ($f_{*,10}^{\rm atom}$, $f_{*,7}^{\rm mol}$ 
and $\alpha_*$) set the normalizations and scaling index.

We assume that the stellar mass is on average built-up over some fraction of the Hubble time, $t_\ast H(z)^{-1}$, resulting in a star formation rate of SFR = $M_\ast H(z)/ t_\ast$.  Here, for computational convenience, we fix $t_\ast = 0.5$ (corresponding to $\sim$ few times the halo dynamical time), 
noting that there is a strong degeneracy between $f_\ast$ and $t_\ast$, and the prior distribution over these two parameters results in a relative insensitivity of the results to $t_\ast$ (\citealt{Park2019MNRAS.484..933P}; \citetalias{Qin2020arXiv200304442Q}).  
To compare with observed UV LFs (e.g. \citealt{Finkelstein2015ApJ...810...71F,Bouwens2015a,Bouwens2016,Livermore2017ApJ...835..113L,Atek2018MNRAS.479.5184A,Oesch2018ApJ...855..105O,Bhatawdekar2019MNRAS.tmp..843B}), we also compute the corresponding $1500{\rm \AA}$ luminosity with a conversion factor ${L_{1500}}/{\rm SFR}=8.7\times10^{27}{\rm erg\ s^{-1}Hz^{-1}yr}$ \citep{Madau2014ARA&A..52..415M}.

We allow ACGs and MCGs to have different UV ionizing 
escape fractions, also with power law scalings with halo mass.  However, for computational convenience, here we assume no evolution with halo mass or redshift resulting in just two additional free parameters, $f_{\rm esc}^{\rm atom}$ 
and $f_{\rm esc}^{\rm mol}$).

The dominant sources of X-rays in the very early Universe are expected to be high mass X-ray binaries (HMXBs; \citealt{Sanderbeck_2018}).  Motivated by models and observations of HMXBs (e.g. \citealt{Mineo2012,Fragos2013ApJ...776L..31F,Pacucci2014}), we assume their population-averaged specific X-ray luminosity scales linearly with the SFR of host galaxies, and has a power-law SED with an energy spectral index of -1. We assume that only X-rays with energy greater than $E_0=$ 500 eV can escape the host galaxy and interact with the IGM.  This value is motivated by high resolution hydrodynamic simulations of the ISM in the first galaxies \citep{Das_2017}. Moreover, we characterize the X-ray luminosity of early galaxies with their {\it soft-band} ($<$2keV) X-ray luminosities, as harder photons have a mean free path longer than the Hubble length and thus do not interact with the IGM. In other words, the specific X-ray luminosity is described by
\begin{equation}
\frac{{{\rm d}L_{\rm X/\dot{\odot}}}}{{\rm d}E} =
\frac{E^{-1}}{\int_{{\rm 
500eV}}^{\rm 2keV}{\rm d}EE^{-1}}\times
	\begin{cases}
	L_{\rm X<2keV/\dot{\odot}}^{\rm atom}\\
	L_{\rm X<2keV/\dot{\odot}}^{\rm mol}
	\end{cases},
\end{equation}
where we include two more free parameters (i.e. $L_{\rm 
X<2keV/\dot{\odot}}^{\rm atom}$ and $L_{\rm X<2keV/\dot{\odot}}^{\rm mol}$) as 
the total soft-band luminosity per SFR for ACGs and MCGs. 

Based on these galaxy properties, we can calculate 1) the ionization and heating rates by X-rays;
2) the {\lya} coupling coefficient between the IGM spin and kinetic temperatures;
3) the LW radiation intensity and the critical halo mass characterising 
the radiative feedback from LW suppression; as well as
4) the UV ionizing photon budget and the critical mass for 
photoheating feedback (see equation \ref{eq:hmf}).
It is worth noting that we also include inhomogeneous recombinations \citep{Sobacchi2014MNRAS.440.1662S}, adopting a sub-grid density distribution from \citet{Miralda2000ApJ...530....1M} but adjusted for the mean density in each cell, and account for density-dependent attenuation of the local ionizing background according to \citep{Rahmati2013MNRAS.430.2427R}.
These quantities are then used to follow the temperature and ionization state of each gas element in our simulation, which are in turn used to compute the 21-cm signal.
For more details see \citet{Mesinger2011MNRAS.411..955M} and \citetalias{Qin2020arXiv200304442Q}.

\section{Mock 21-cm observation}\label{sec:mock}

\begin{table*}
	\caption{A list of the free parameters varied in this work, together with their descriptions, fiducial values used for the mock observation, and the
        recovered 
values (median with [14, 86] percentiles) obtained from the two {\tocmmc} 
runs. Other model parameters are held constant, using the values discussed in the text and in
\citetalias{Qin2020arXiv200304442Q}.}
	\vspace{-3mm}
	\begin{threeparttable}
		\label{tab:parameters}
		\begin{tabular}{llccc}
			\hline \hline
			\multirow{2}{*}{Parameters} & 
\multirow{2}{*}{Description} & \multirow{2}{*}{Mock} 
&\multicolumn{2}{c}{{\tocmmc} results}\\\cline{4-5}
			&&&{\it 2pop}& {\it 1pop}\\\hline
			$\log_{10}f_{*,10}^{\rm atom}$ & 
\multirow{2}{*}{Stellar to halo mass ratio} \multirow{2}{*}{at 
				$M_{\rm vir}=\begin{array}{l}10^{10}\msol \\ 
10^{7}\msol \end{array}$for$\begin{array}{l}\rm ACGs \\ \rm MCGs\end{array}$}& 
$-1.25$&$-1.28^{+0.05}_{-0.19}$&$-1.28^{+0.06}_{-0.24}$\\
			\vspace{1mm}
			$\log_{10}f_{*,7}^{\rm mol}$ &&$-2.75$ 
&$-3.08^{+0.89}_{-0.61}$ &-\\\hdashline
			\vspace{1mm}
			$\alpha_*$ & Stellar to halo mass power-law  index& 
$0.5$&$0.52^{+0.16}_{-0.11}$&$0.43^{+0.27}_{-0.09}$\\\hdashline
			\vspace{1mm}
			$\log_{10}f_{\rm esc}^{\rm atom}$ & 
\multirow{2}{*}{Escape fraction of ionizing photons } 
\multirow{2}{*}{for$\begin{array}{l}\rm ACGs \\ \rm MCGs\end{array}$}&  
$-1.22$& $-1.23^{+0.14}_{-0.21}$&$-1.14^{+0.12}_{-0.14}$\\
			\vspace{1mm}
			$\log_{10}f_{\rm esc}^{\rm mol}$ && 
$-1.22$&$-1.38^{+0.93}_{-1.05}$&-\\\hdashline
			\vspace{1mm}
			$\log_{10}L_{\rm X<2keV/\dot{\odot}}^{\rm atom}$ & 
\multirow{2}{*}{Soft-band X-ray luminosity per SFR ($\rm erg\ s^{-1} \msol^{-1} 
yr$)} \multirow{2}{*}{for$\begin{array}{l}\rm ACGs \\ \rm MCGs\end{array}$}& 
40.5&$40.64^{+1.28}_{-1.88}$& $41.21^{+1.17}_{-0.22}$ \\
			\vspace{1mm}
			$\log_{10}L_{\rm X<2keV/\dot{\odot}}^{\rm mol}$ 
&&41.7&$41.77^{+1.07}_{-2.01}$&-\\	
			\hline 
			\hline
		\end{tabular}
	\end{threeparttable}
\end{table*}

We create a mock 21-cm observation from a simulation box with a comoving volume of $(500{\rm Mpc})^3$ and a 256$^3$ grid.
While the full 
parameter space of our model is very large (17 dimensional; see Table 1 in 
\citetalias{Qin2020arXiv200304442Q}), in this proof-of-concept work, we limit it to the 7 parameters that drive the largest signal variation and are most relevant for the early Cosmic Dawn signal.  These include
$f_{*,10}^{\rm atom}$, $f_{*,10}^{\rm mol}$, $\alpha_*$, 
$f_{\rm esc}^{\rm atom}$, $f_{\rm esc}^{\rm mol}$, $L_{\rm 
X<2keV/\dot{\odot}}^{\rm atom}$ and $L_{\rm X<2keV/\dot{\odot}}^{\rm mol}$. 
Table \ref{tab:parameters} summarizes their physical meaning, and shows the fiducial values we use to make our mock observation.  These fiducial values are chosen in order for the mock observation to be consistent with the following observations:
\begin{enumerate}
	\item the galaxy UV LFs at $z\sim6{-}10$ 
\citep{Bouwens2015a,Bouwens2016,Oesch2018ApJ...855..105O};
	\item the upper limit on the neutral fraction at $z{\sim}5.9$ from the dark fraction in QSO spectra 
($x_{\hone}<0.06{+}0.05, 1\sigma$; \citealt{McGreer2015MNRAS.447..499M});
	\item the CMB Thomson scattering optical depth from {\it 
Planck} ($\tau_e=0.058{\pm}0.012, 1\sigma$; 
\citealt{Planck2016A&A...596A.108P}); and
	\item the timing\footnote{We only consider 
the timing from EDGES that is expected to be driven by minihalos (see 
\citetalias{Qin2020arXiv200304442Q} and also 
\citealt{Mirocha2019MNRAS.483.1980M}). This allows us to select an optimistic model for our proof-of-concept study, in which minihalos play an important role.  The amplitude of the reported signal cannot be explained by standard physics (e.g. \citealt{Munoz2018Natur.557..684M,Fialkov2018PhRvL.121a1101F,Ewall-Wice2018ApJ...868...63E,Mebane2019arXiv191010171M}), and some exotic explanations could have a large impact also on the power spectrum. However, {\it partial} degeneracy with unidentified systematics and/or foregrounds (e.g. \citealt{Hill2018JCAP...08..037H,Spinelli2018MNRAS.479..275S,Bradley2019ApJ...874..153B,Sims2019MNRAS.488.2904S})
could bring the amplitude in line with standard models, without evoking exotic physics.  Our mock PS corresponds to such a scenario.}
of the recent putative detection of an absorption profile centered at $78\pm1$MHz in the global 21-cm spectrum by EDGES \citep{Bowman2018}.
\end{enumerate}

We note that, $f_{*,10}^{\rm atom}{\sim}6\%$ and $\alpha_*{=}0.5$ are already
well constrained by the observed high-redshift UV LFs while $f_{\rm esc}^{\rm 
atom}{=}6\%$
ensures reionization of the fiducial model finishes by $z{\sim}5.9$, with the inferred 
$\tau_e$ consistent with results from {\it Planck}. On the other hand, 
$\log_{10}\left[L_{\rm X<2keV/\dot{\odot}}^{\rm atom}/{\rm erg\ 
s^{-1}{\msol}^{-1}yr}\right]{=}40.5$ is motivated by
theoretical models of high-mass X-ray binaries in metal-poor environments
\citep{Fragos2013ApJ...776L..31F}. Without much knowledge of the MCG 
properties, we assume that their stellar to halo mass relation and escape 
fraction follow ACGs, and choose $f_{*,7}^{\rm mol}{\sim}0.2\%$ and $f_{\rm 
esc}^{\rm mol}{=}6\%$. Finally, an enhanced X-ray luminosity of MCGs, here we take 
$\log_{10}\left[L_{\rm X<2keV/\dot{\odot}}^{\rm atom}/{\rm erg\ 
s^{-1}{\msol}^{-1}yr}\right]{=}41.7$, is needed\footnote{This is based on the 
strong degeneracy between $f_{*,7}^{\rm mol}$ and the X-ray luminosity per SFR 
found in \citetalias{Qin2020arXiv200304442Q}. Note that although ACGs and MCGs 
are assumed to share the same specific X-ray luminosity in 
\citetalias{Qin2020arXiv200304442Q}, a smaller $L_{\rm X<2keV/\dot{\odot}}^{\rm 
atom}$ used in this work does not have a significant impact to the timing of the absorption 
trough because the contribution from ACGs at high redshifts ($z{\gtrsim}15$) is 
small.} to reproduce an 21-cm absorption trough centred at ${\sim}78$MHz 
(\citetalias{Qin2020arXiv200304442Q}).  This could be motivated by more luminous X-ray binaries arising from Pop-III stellar remnants in MCGs \citep{Xu2016ApJ...832L...5X}.  It is important to note that these are just fiducial parameter values, chosen to make the mock observation consistent with our current knowledge; there are large uncertainties and strong degeneracies as we will see below.

We present the 21-cm lightcone from our fiducial model in the upper panel 
of Fig. \ref{fig:MCMC_PS} and show its globally averaged 21-cm brightness 
temperature evolution, EoR history as well as the Thomson scattering 
optical depth ($\tau_e=0.062$) in the upper right three sub-panels of Fig. 
\ref{fig:MCMC} using black solid curves. We see that the model is consistent with the aforementioned 
observational constraints.
The signal follows the expected qualitative trends (e.g. \citealt{Furlanetto2006PhR...433..181F,Baek2010A&A...523A...4B,Santos2011A&A...527A..93S,Mesinger2016,Park2019MNRAS.484..933P}).
During the cosmic dawn, the first galaxies 
begin to build up the {\lya} background, coupling the spin ($T_{\rm 
s}$) and kinetic temperatures ($T_{\rm k}$). The brightness 
temperature ($\delta T_{\rm b}$) is negative (i.e. the IGM is seen in absorption against the CMB) and decreases as the IGM adiabatically cools faster than the CMB. For our choice of galaxy parameters, $\delta T_{\rm 
b}$ reaches its minimum at $z{\sim}17$, before X-ray 
heating becomes significant, eventually heating the IGM to temperatures above the CMB by $z{\sim}13-14$.
As reionization progresses, the signal starts fading 
until $z\sim{5.8}$ when the universe is fully ionized (apart from the residual $\hone$).

\subsection{Synthetic power spectra and telescope noise}\label{subsec:mock_ps}

\begin{figure*}
	\begin{minipage}{\textwidth}
		\begin{center}
			\includegraphics[width=\textwidth]{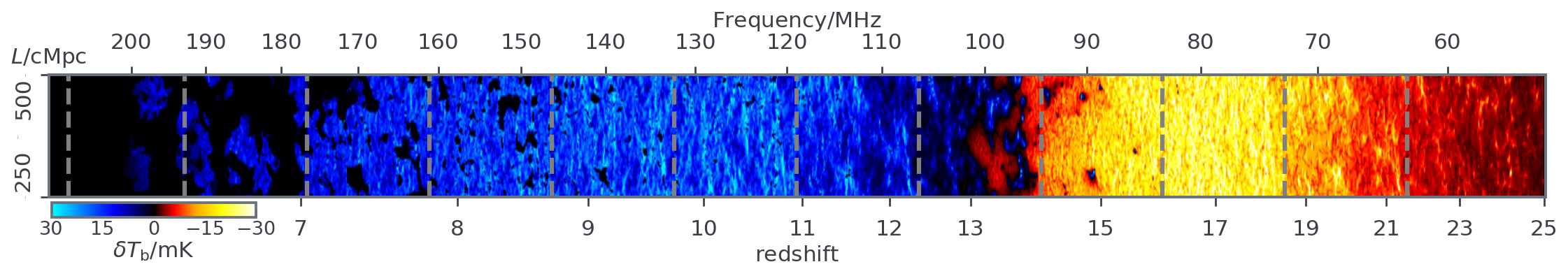}\\
			\includegraphics[width=\textwidth]{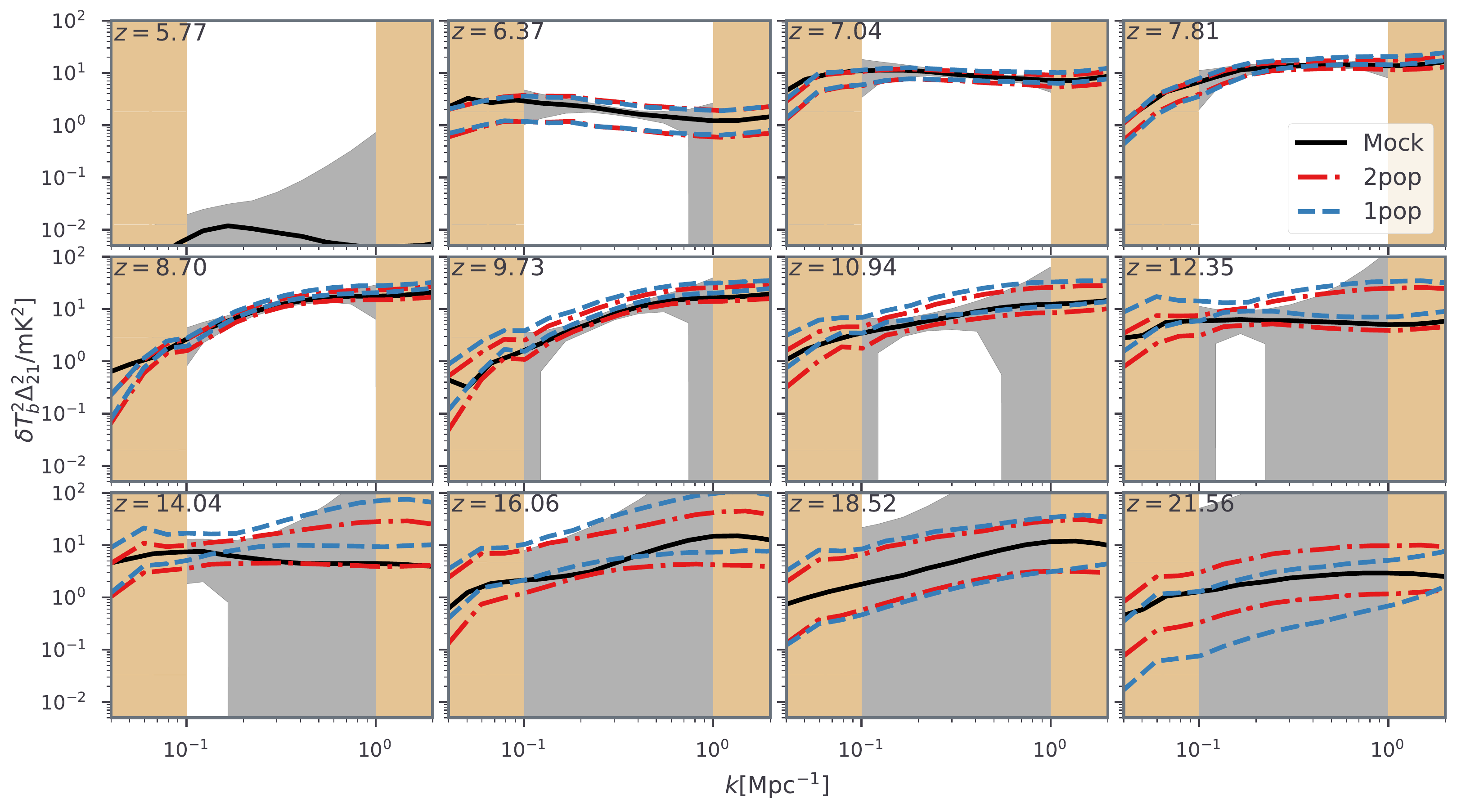}
		\end{center}
	\end{minipage}
	\caption{\label{fig:MCMC_PS}{\it Top panel}: A slice of the 21-cm 
		lightcone from our mock observation. The central redshifts of the 12 
		independent box samples, which are used to calculate the 21-cm PS, are 
		indicated by the vertical dashed lines. Note that the spatial range of the vertical axis is from 250 to
		500cMpc, half the entire lightcone length (500cMpc).
		{\it Lower panels:} evolution of the 
		21-cm PS.  Solid black curves correspond to the mock observation, with gray shaded regions indicating the 1$\sigma$ noise from a 1000h observation with SKA1-low.  Only power 
		within the range of $k=0.1-1{\rm Mpc}^{-1}$ is considered when performing the 
		Bayesian inference.
        The [14, 86] percentiles of the recovered  posteriors from Fig. \ref{fig:MCMC}
                are bracketed by the colored lines ({\it 2pop} using red dash-dotted lines, {\it 1pop} using blue dashed lines).}
\end{figure*}

Following \citet{Greig2018MNRAS.477.3217G}, we compress the cosmic 21-cm lightcone into 3D averaged power spectra. The PS from the mock observation (generated from a unique initial seed) and the forward-modelled simulations (in Sec. \ref{sec:constraint}) are calculated from the same comoving volume of the lightcone.
For computing efficiency, forward-modeled simulations have a factor of $2^3$ smaller volume than the mock while keeping the same resolution, i.e. $(250{\rm Mpc})^3$ and 128$^3$ cells. Therefore, we compute the PS from 12 independent sub volumes of the lightcone between $z=5.5$ (${\sim}220$MHz) and $z=30$ (${\sim}50$MHz).  The resulting 3D-averaged PS are shown with the black curves in the sub-panels of Fig. \ref{fig:MCMC_PS}, with the central redshifts indicated in the top left of each panel and with the vertical lines in the top panel.

We add to our cosmic 21cm PS instrument noise corresponding to a 1000h integration with SKA1-low\footnote{Note that HERA is expected to provide comparable astrophysical parameter recovery as SKA1-low, using the power spectrum summary statistic and the fiducial galaxy models from \citet{Park2019MNRAS.484..933P} and \citet{Greig2020MNRAS.491.1398G}.
However, our fiducial model here is chosen to have a significant contribution of MCGs driving a much earlier epoch of heating, as motivated by the putative EDGES detection.  As thermal noise dominates at the corresponding low frequencies and SKA1-low should have smaller thermal noise than HERA for a fixed integration time, for this proof-of-concept study we make the slightly more optimistic choice of SKA1-low when computing the noise.}.
We use the public \textsc{21cmsense} package \citep{Pober2013AJ....145...65P,Pober2014ApJ...782...66P}. 
We adopt the ``moderate'' foreground removal configuration, which excises foreground contaminated modes from the cylindrical $k$-space ``wedge'' (which is assumed to extend at $k_{\parallel} \approx 0.1 h {\rm Mpc}^-1$ above the horizon limit) and assumes coherent addition of only instantaneously redundant baselines (see more in  \citealt{Pober2014ApJ...782...66P}).
We provide a brief summary of the relevant calculations here and direct interested readers to the aforementioned papers for more details.

The thermal noise PS of a single baseline corresponding to a given $k$ mode is \citep{Morales2005ApJ...619..678M,McQuinn2006ApJ...653..815M,Parsons2012ApJ...753...81P}
\begin{equation}
	\Delta_{\rm N}^2(k) \approx \frac{3k^3(1+z)^4}{2\pi^2\sqrt{\Omega_{\mathrm{m}}(1+z)^3+\Omega_{\mathrm{\Lambda}}}}\Omega B T_{\rm N}^2,
\end{equation}
where $\Omega$ and $B$ correspond to the solid angle of the primary beam size (e.g. ${\sim}0.1$sr at $z=17$) and observing bandwidth (8MHz), respectively.
We use the SKA1-low antennae configuration described in \citet{Greig2020MNRAS.491.1398G}.
The system temperature is taken to be $T_{\rm N}=(T_{\rm sky}+T_{\rm rec}) (2Bt)^{-0.5}$ where $T_{\rm sky}$ and $T_{\rm rec}$ represent the sky and receiver temperatures while the factor, $\sqrt{2Bt}$, reflects the number of independent measurements during the integration time, $t$.  Following \citet{Thompson2017isra.book.....T}, the sky is modelled as being dominated by Galactic synchrotron emission and scales with frequency ($\nu$) as $T_{\rm sky}=60{\rm K}(\nu/300{\rm MHz})^{-2.55}$. On the other hand, the receiver is assumed to be kept at 40K with an addition of $0.1T_{\rm sky}$ reflecting its response to the sky \citep{Pober2014ApJ...782...66P}. 

The total uncertainty on the 21-cm PS ($\sigma \Delta_{21}^2$) is obtained by summing over the individual modes, $i$, \citep{Pober2013AJ....145...65P}, and adding the cosmic variance of the mock observation (reasonably assuming it is Gaussian distributed at the relevant scales; \citealt{Mondal2016MNRAS.456.1936M}:
\begin{equation}\label{eq:sigmaPS}
	\left[\frac{1}{\sigma \Delta_{21}^2 (k)}\right]^2 = \sum_{i}\left(\frac{1}{\Delta_{\rm N,i}^2+\Delta_{21}^2}\right)^2.
\end{equation} 

The gray shaded regions in Fig. \ref{fig:MCMC_PS} show the resulting 1$\sigma$ uncertainty on the mock cosmic signal.  We note large uncertainties at high redshifts while most constraints from the 21-cm PS come from large scales and $z<15$.  When performing inference, we additionally exclude the modes outside the range  $0.1\le k/{\rm Mpc^{-1}} \le 1$ (demarked in brown in the panels).  This is done to conservatively avoid additional foreground contamination as well as aliasing (shot noise) effects from our simulation grids.  Moreover, we add an additional 20\% ``modeling error'' to our forward-modeled PS, roughly motivated by comparisons to radiative transfer simulations (e.g. \citealt{Zahn2011MNRAS.414..727Z}).  We note however that such modeling error is unlikely to have a major impact on parameter inference \citep{Greig2020MNRAS.491.1398G}.

\section{Can we indirectly detect the first, molecularly-cooled galaxies?}
\label{sec:constraint}

In this section, we use {\tocmmc} \citep{Greig2015MNRAS.449.4246G} to constrain astrophysical parameters and perform model selection using the following observations:
\begin{itemize}
\item the mock 21-cm power spectra discussed in Sec. \ref{subsec:mock_ps};
\item the observed galaxy LFs at $z{\sim}6{-}10$ \citep{Bouwens2015a,Bouwens2016,Oesch2018ApJ...855..105O};
\item the upper limits on the neutral fraction at $z{\sim}5.9$ from QSO spectra \citep{McGreer2015MNRAS.447..499M}; and
\item the Thomson scattering optical depth of the CMB \citep{Planck2016A&A...596A.108P}.
\end{itemize}
We perform inference using the following two models:
\begin{enumerate}
\item {\it 2pop}: the ``full'' model, including both MCGs and ACGs, used to generate the mock observation; and
\item {\it 1pop}: a single population model consisting only of ACGs.
\end{enumerate}
{\it 2pop} is characterized with the 7 free parameters listed in Table \ref{tab:parameters}, while {\it 1pop} only has the four parameters relevant for ACGs (i.e. excluding the ones labelled ``mol'').  It is clear that {\it 1pop} cannot fully reproduce the mock observation: ACGs are too biased at early times and are not sensitive to the build up of the inhomogeneous LW background.  However, given the limited sensitivity of even SKA1-low during the Cosmic Dawn, will we be able to say with certainty that the {\it 1pop} model is incorrect?

This question can be readily answered with Bayesian inference.
Using the built-in Occam's razor in the Bayes factor, we can quantify whether the unique properties of MCGs are needed to explain the ``observation'', or whether the simpler, single-population model can adequately mimic the signal.  Is the additional model complexity of {\it 2pop} justified by the data?  If not, we might not be able to detect minihalo galaxies even indirectly  with upcoming interferometers. 

Before presenting our results in Sec. \ref{subsec:results}, we briefly review the basics of Bayesian model selection (Sec. \ref{subsec:bayes}), as well as the \textsc{MultiNest} sampler we use inside {\tocmmc} (Sec. \ref{subsec:multinest}).

\subsection{Bayesian evidence and model selection}\label{subsec:bayes}

Bayes' law states that the posterior probability distribution 
[$P(\theta|\mathscr{O},\mathscr{M})$] of model ($\mathscr{M}$) characterized by parameters 
($\theta$) when constrained by observations ($\mathscr{O}$) is equal to 
the product of our prior knowledge  [$P(\theta|\mathscr{M})$] and the likelihood 
function [$P(\mathscr{O}|\theta, \mathscr{M})$] divided by the evidence
[$P(\mathscr{O}|\mathscr{M})$]
\begin{equation}\label{eq:bayes}
P(\theta|\mathscr{O},\mathscr{M})= P(\theta|\mathscr{M})
\frac{P(\mathscr{O}|\theta, \mathscr{M})}{P(\mathscr{O}|\mathscr{M})}
.
\end{equation}

While the posterior represents our belief about the model after taking the observation 
into account, the prior reflects our knowledge before\footnote{Here we use a flat prior over the following ranges:
$f_{*,10}^{\rm atom}\in[10^{-3},1]$ in logarithmic space; $f_{*,10}^{\rm mol}\in[10^{-4},10^{-1}]$ in logarithmic space; $\alpha_*\in[-0.5,1]$; 
$f_{\rm esc}^{\rm atom(mol)}\in[10^{-3},1]$ in logarithmic space; and $L_{\rm X<2keV/\dot{\odot}}^{\rm atom(mol)}{\in}[10^{38},10^{44}]\ {\rm erg\ s^{-1} \msol^{-1}yr}$ in logarithmic space.}.
The likelihood  measures how well a parameter combination $\theta$ can reproduce the observed data $\mathscr{O}$.

The Bayesian evidence, also known as the marginal likelihood, is central to model selection.
It can be 
computed by integrating the likelihood, weighted by 
the prior, over the entire parameter space:
\begin{equation}\label{eq:evidence}
\begin{split}
P(\mathscr{O}|\mathscr{M}) & = \oint_{\theta} {\rm d}\theta 
P(\mathscr{O}|\theta, \mathscr{M})P(\theta|\mathscr{M})\\
&{\sim} \delta \theta P(\mathscr{O}|\theta_{\rm max}, \mathscr{M})P(\theta_{\rm max}|\mathscr{M})
\end{split}
\end{equation}
The last step in equation (\ref{eq:evidence}) approximates the integral trapezoidally around the maximum likelihood, $P(\mathscr{O}|\theta_{\rm max}, \mathscr{M})$ (e.g. \citealt{Trotta2017arXiv170101467T}). Here, $\delta \theta$ and $\Delta \theta {\equiv} P^{-1}(\theta_{\rm max}|\mathscr{M})$ characterize widths of the likelihood and prior, respectively. 
The factor
$\delta \theta/\Delta \theta$
is commonly referred to as Occam's factor, as it penalizes models which have a prior volume that is larger than  the likelihood.

There are many model selection criteria
\citep{Liddle2004MNRAS.351L..49L,Liddle2007MNRAS.377L..74L}
to answer whether the increased complexity (see more in 
\citealt{Kunz2006PhRvD..74b3503K}) of a model involving a
higher-dimensional parameter space
is justifiable by the observation -- in our case,
whether upcoming 21-cm PS measurements can be used to detect minihalo-hosted galaxies. Here. we use an empirical scale \citep{Jeffreys1939thpr.book.....J} based on the ratio of the evidences of the two models, the so-called Bayes factor.  Specifically, the
 probability of the {\it2pop} model
being preferred over {\it1pop} is 75.0\% (weak), 92.3\% (moderate) and 99.3\% (strong) if 
$\ln B \equiv\ln\left[P(\mathscr{O}|\mathscr{M}_{\it 2pop})/P(\mathscr{O}|\mathscr{M}_{\it 1pop})\right]$
is 1, 2.5 and 5, respectively.

\subsection{Including \textsc{MultiNest} in 21CMMC}\label{subsec:multinest}
\begin{figure*}
	\begin{minipage}{\textwidth}
		\begin{center}
			\includegraphics[width=\textwidth]{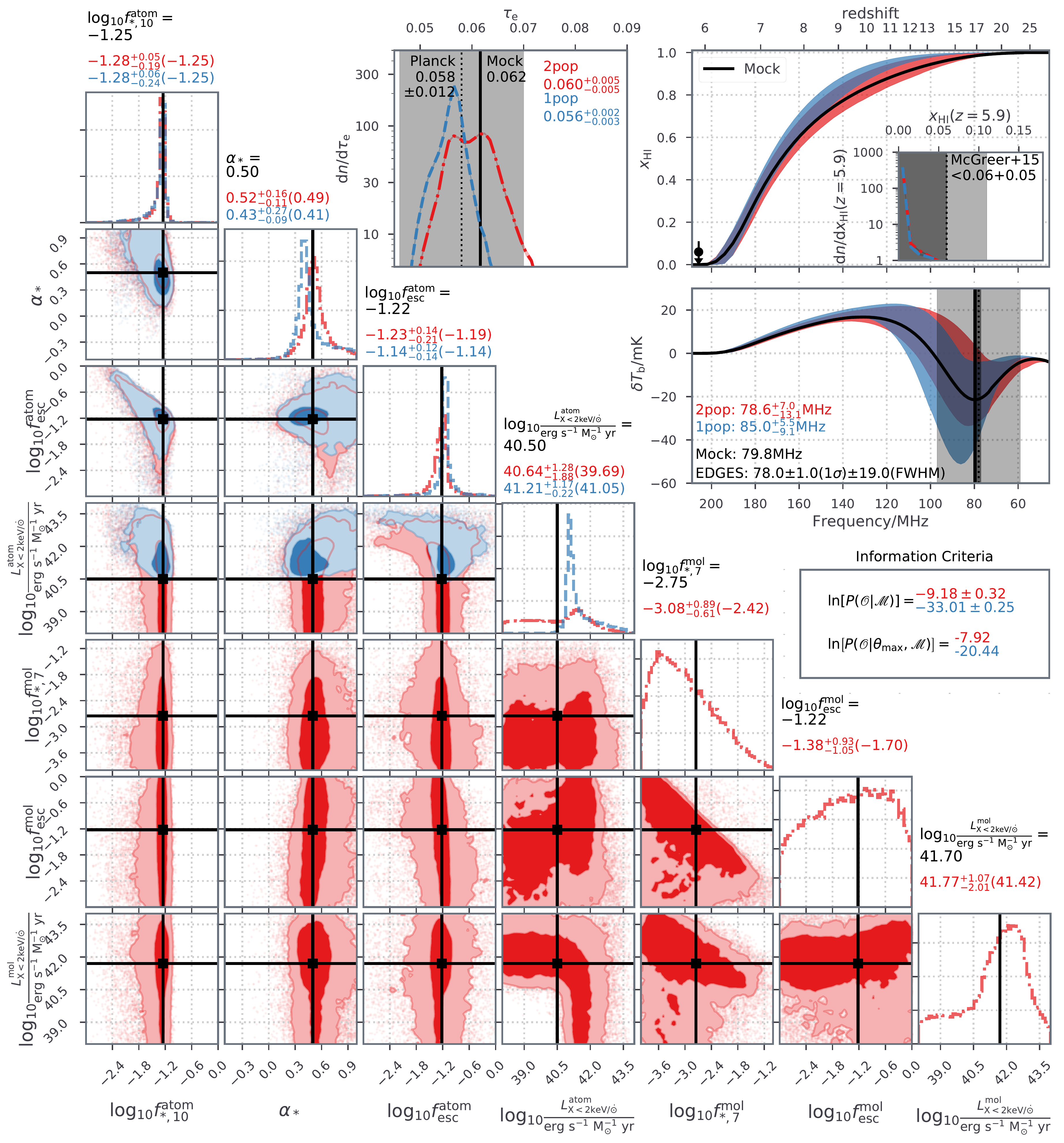}\vspace*{2mm}
		\end{center}
	\end{minipage}
\caption{\label{fig:MCMC} Marginalized posterior distributions from our two astrophysical models: (i) {\it 2pop} in red / dash-dotted lines; (ii) {\it 1pop} in blue / dashed lines. While the {\it 1pop} model only considers 4 parameters describing ACGs,
                {\it 2pop} includes additional 3 parameters representing the properties of MCGs.
                Both results use the following observations when computing the likelihood:
                (i) the observed galaxy LFs at $z{\sim}6{-}10$ \citep{Bouwens2015a,Bouwens2016,Oesch2018ApJ...855..105O};
                (ii) the upper limits on the neutral fraction at $z{\sim}5.9$ from QSO spectra 
		\citep{McGreer2015MNRAS.447..499M};
                (iii) the Thomson scattering optical depth of the CMB 
		\citep{Planck2016A&A...596A.108P}; and
                (iv) the 21-cm power spectra from a mock observation (chosen to be consistent with timing implied by the putative EDGES detection; \citealt{Bowman2018}; see the black solid lines in Fig. \ref{fig:MCMC_PS}).
                The 2D distributions correspond to 68th (dark regions) and 95th (light regions) percentiles. 
		The medians with [14, 86] percentiles for each parameter 
		are presented on the top of the 1D PDF together with the values for the maximum likelihood models (in brackets). 
		The mock parameters are indicated by solid black lines in the posterior with their values shown on the top of the 1D PDFs as well.
		The upper right three sub-panels present the [14, 86]
		percentiles of the volume weighted neutral hydrogen fraction ($x_{\hone}$) and brightness temperature 
		($\delta T_{\rm b}$) as well as the PDF of $\tau_e$ and 
		$x_{\hone}$ at $z{=}5.9$
		for the two  posterior 
		distributions. Fiducial values from the mock observation are denoted with solid black curves.
                Observations are indicated in grey or using black circles. 
}
\end{figure*}

{\tocmmc}\footnote{\url{https://github.com/21cmfast/21CMMC}} \citep{Greig2015MNRAS.449.4246G} is a Bayesian sampler of 21-cm lightcones, allowing for cosmological and astrophysical parameter inference from the 21-cm signal \citep{Greig2017MNRAS.472.2651G,Greig2018MNRAS.477.3217G}.
In its default configuration,
{\tocmmc} employs an ensemble sampler (\textsc{emcee};
\citealt{Goodman2010CAMCS...5...65G,Foreman2013PASP..125..306F,Akeret2013A&C.....2...27A})
to explore the parameter space, which does not require the evidence to generate a proposal distribution.  This makes the evaluation of the Bayesian evidence computationally challenging in a high-dimensional parameter space (see the first part of equation \ref{eq:evidence}).

In this work, we include the \textsc{MultiNest}\footnote{\url{https://github.com/rjw57/MultiNest}\\ \url{https://github.com/JohannesBuchner/PyMultiNest}} sampler \citep{Feroz_2008,Feroz_2009,Feroz_2019,Buchner2014A&A...564A.125B} in {\tocmmc},
which implements nested sampling -- 
converting the variable of integration in equation (\ref{eq:evidence}) from the high-dimensional parameter space to the 1D prior space (see more in \citealt{Skillingdoi:10.1063/1.1835238})
\begin{equation}\label{eq:nested}
P(\mathscr{O}|\mathscr{M}) =\int_{0}^{1} {\rm d}P(\theta|\mathscr{M}) 
P[\mathscr{O}|P(\theta|\mathscr{M})],
\end{equation}
where ${\rm d}P(\theta|\mathscr{M}) \equiv d\theta P(\theta|\mathscr{M})$ represents the differential of prior volume. By reducing the prior volume around higher probability regions
at each step when new sampling points are drawn, 
\textsc{MultiNest} computes the posterior and calculates the Bayesian evidence as a ``by-product". The current public version of {\tocmmc} allows the user to choose between \textsc{emcee} and \textsc{MultiNest} samplers.

It is worth noting that
the recent development of \textsc{21cmNest} by \citet{Binnie_2019} also introduced 
\textsc{MultiNest} into {\tocmmc}. They found the posterior of a 3-parameter
21-cm model inferred from mock observations to be consistent between \textsc{21cmNest}
and the original {\tocmmc}. This encourages us to apply it to our updated 21-cm simulations using 
more sophisticated galaxy models.

\subsection{Strong evidence of minihalos}\label{subsec:results}

In Fig. \ref{fig:MCMC}, we present the marginalized posteriors from our two models ({\it 1pop}/{\it 2pop} in blue/red), including model parameters, global 21-cm signals, EoR histories, and the optical depths.  The corresponding 21-cm power spectra are shown in Fig. \ref{fig:MCMC_PS}.

For both models, the properties of ACGs are tightly constrained,
including $f_{*,10}^{\rm atom}$, $\alpha_*$ and $f_{\rm esc}^{\rm atom}$.
We caution however that these parameters, especially $f_{\rm esc}^{\rm atom}$, are {\it overconstrained} (e.g. compared to \citealt{Park2019MNRAS.484..933P}) due to the fact that several ACG parameters are kept fixed in this demonstrative study (most importantly $M^{\rm SN}_{\rm crit}$, $\alpha_{\rm esc}$, $t_\ast$).

In the absence of MCGs, we see that the {\it 1pop} model dramatically overestimates the X-ray luminosities of ACGs, with the 1D PDF peaking sharply at $\log_{10}(L_{\rm X<2keV/\dot{\odot}}^{\rm atom}/\rm erg\ s^{-1} \msol^{-1} yr){\sim}41.0-42.4$: a factor of ${\sim}3-75$ times higher than the ``true'' value of the mock signal.  Moreover, the {\it 1pop} model prefers a lower $\alpha_\ast$ (i.e. a steeper stellar mass function), despite the fact that the UV LFs already constrain this parameter (e.g. \citealt{Park2019MNRAS.484..933P}).  Thus, the ${\it 1pop}$ posterior prefers galaxy models with more efficient star formation in lower mass halos (i.e. smaller $\alpha_*$) and with higher X-ray emissivities (i.e. larger $L_{\rm X<2keV/\dot{\odot}}^{\rm atom}$), in order to (partially) compensate for the missing population of MCGs.

From the global evolution of the neutral fraction and brightness temperatures, as well as the power spectra, we see that the {\it 1pop} model does indeed perform a reasonable job at capturing the mock observation.  Differences emerge at the highest redshifts, when the radiation fields have a higher relative contribution from MCGs.  Even with a higher X-ray emissivity and steeper stellar mass functions, the ACG-only model cannot fully capture the evolution of the ACG + MCG mock observation.  ACGs are more biased galaxies, and are insensitive to LW feedback which can prolong the early evolution of IGM properties in feedback-dominated MCG models (e.g. \citealt{Ahn2009ApJ...695.1430A,Holzbauer2012MNRAS.419..718H,Fialkov2013MNRAS.432.2909F}).  Thus, compared to the mock signal, the {\it 1pop} model has: (i) a more rapid evolution of cosmic milestones; and (ii) a higher 21-cm PS during the epochs when a single field (i.e. temperature or Ly$\alpha$ coupling) sources the fluctuations, thus making cross terms negligible and allowing the 21-cm PS to be roughly estimated analogously to the halo model with a bias term for the galaxies (e.g. \citealt{Pritchard2007MNRAS.376.1680P,McQuinn_2018}).
Indeed, we see that the {\it 1pop} model has a more rapid evolution of the early stages of reionization (see also \citealt{Ahn2009ApJ...695.1430A}).  Moreover, during the epoch of heating when the 21-cm signal is sourced by temperature fluctuations ($12 \lsim z \lsim 15$), {\it 1pop} prefers power spectra that are too high, and results in a too rapid evolution during the transition to the earlier, Ly$\alpha$-dominated epoch ($z\gsim15$).

On the other hand, the ``full'', {\it 2pop} model recovers the fiducial parameters of the mock observation quite well.  The inferred global evolution of the neutral fraction and brightness temperature, as well as the power spectra, are all consistent with the mock observation, without any notable bias.  The X-ray luminosity of MCGs is well constrained, to within $\sim$ 1 dex of the fiducial value.  Interestingly, there is a tail in the PDF extending towards low luminosities.  Looking at the $L_{\rm X<2keV/\dot{\odot}}^{\rm atom}$ -- $L_{\rm X<2keV/\dot{\odot}}^{\rm mol}$ marginalized posterior, we see that this is due to a partial degeneracy allowing ACGs to dominate the epoch of heating for those models in which MCGs do not emit significant soft X-rays.

The ionizing escape fraction of MCGs is poorly constrained, as they do not have a significant contribution to reionization.  However, models with both high $f_{*,7}^{\rm mol}$ and $f_{\rm esc}^{\rm mol}$ are excluded as they would result in a Thomson scattering optical depth that is too high (see \citetalias{Qin2020arXiv200304442Q} and \citealt{Visbal2015MNRAS.453.4456V}).

\begin{figure*}
	\begin{minipage}{\textwidth}
		\begin{center}
			
			\includegraphics[width=0.8\textwidth]{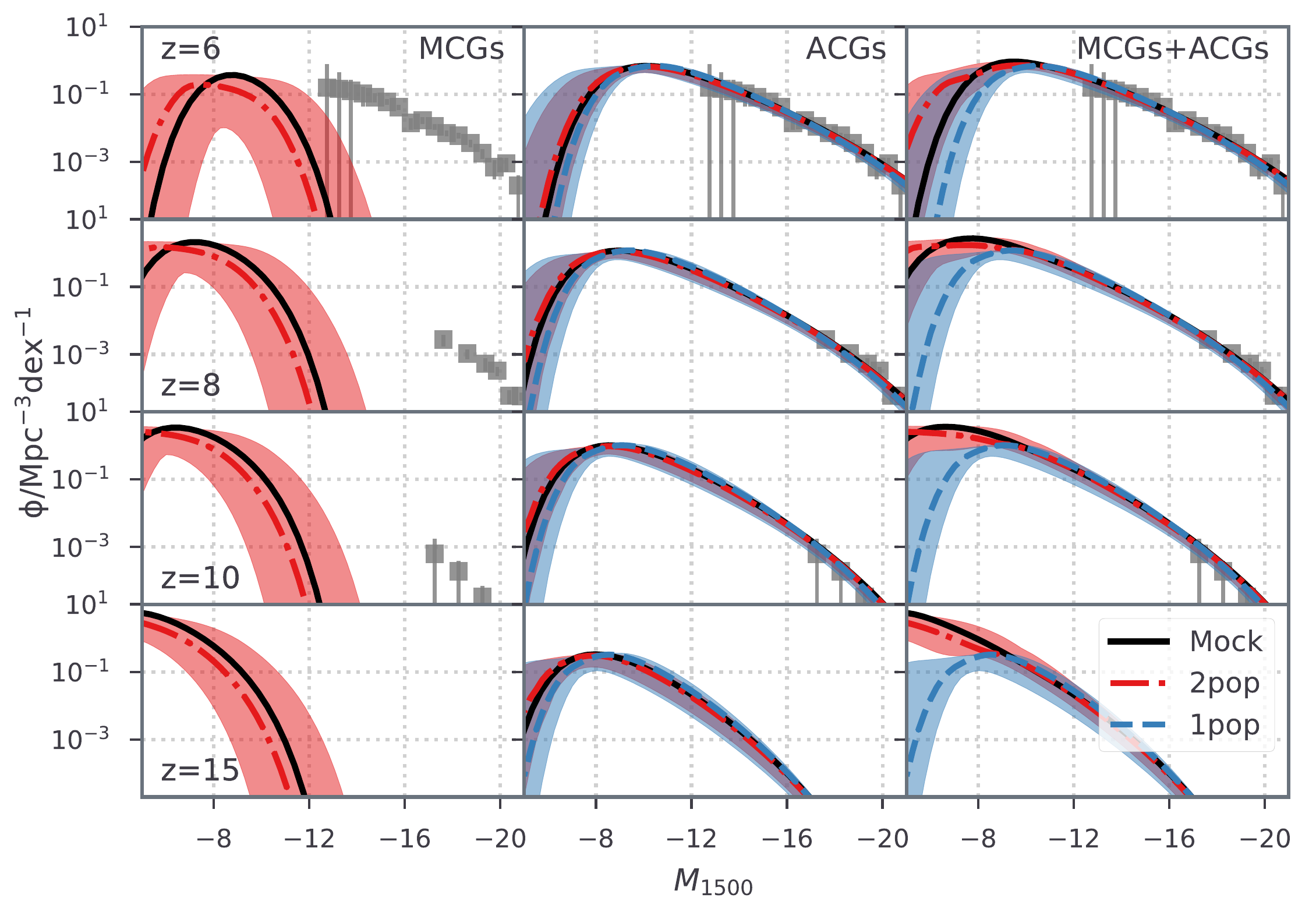}
		\end{center}
	\end{minipage}
	\caption{\label{fig:MCMC_LF}UV luminosity functions of MCGs, ACGs and 
		all galaxies from the model posteriors: ({\it 2pop} in red, {\it 
			1pop} in blue).  Lines and shaded regions represent the median 
		and [16, 84] percentiles. Observational estimates used in the inference
		\citep{Bouwens2015a,Bouwens2017ApJ...843..129B,Oesch2018ApJ...855..105O} are
		shown in grey at the bright end.}
\end{figure*}

For completeness, we also present the marginalized UV LFs of ACGs, MCGs (only in {\it 2pop}) and all galaxies in Fig. \ref{fig:MCMC_LF}. We see that the ACGs and total LFs are tightly constrained at the bright end by currently available observations \citep{Bouwens2015a,Bouwens2016,Oesch2018ApJ...855..105O}. Compared to the mock observation, both {\it 1pop} and {\it 2pop} results are consistent at $M_{1500}<{-}8$. At fainter magnitudes, only the {\it 2pop} model recovers the UV LFs, since MCGs dominate in this regime.

Finally, we come to the main question of this work: can we quantitatively claim that {\it 2pop} is a better fit to the data, given that it has more free parameters compared to {\it 1pop}?  We quantify this using the Bayesian evidence: $\ln\left[P(\mathscr{O}|\mathscr{M})\right]=-9.18\pm0.32$ and $-33.01\pm0.25$
for {\it 2pop} and {\it 1pop}, respectively. These result in a Bayes factor of $\ln B\equiv \ln\left[P(\mathscr{O}|\mathscr{M_{\it 2pop}})/P(\mathscr{O}|\mathscr{M_{\it 1pop}})\right]\sim24$, suggesting the probability of {\it 2pop} being preferred over {\it 1pop} by the data (i.e. the mock 21-cm PS) is ${>}99.3\%$ \citep{Jeffreys1939thpr.book.....J}.
We therefore conclude that the (mock) data require the additional parameters characterizing
MCG  (i.e. $f_{*,7}^{\rm mol}$, $f_{\rm esc}^{\rm mol}$ and $L_{\rm X<2keV/\dot{\odot}}^{\rm mol}$).
This means that it might be possible to indirectly detect the footprint of MCGs in upcoming 21-cm power spectra measurements.
We caution that this conclusion is based on the assumption that minihalo-hosted galaxies truly play a significant role in the IGM evolution during the cosmic dawn (as would be the case if, for example, the EDGES detection is genuinely cosmological).
 
\section{Conclusions}\label{sec:conclusion}
In this work, we quantify the detectability of minihalos for upcoming
21-cm interferometers. We compute a mock 21-cm signal, motivated by the timing of the putative EDGES detection, which would be driven by X-ray luminous, molecularly-cooled galaxies (\citetalias{Qin2020arXiv200304442Q}).  
The result additionally agrees with the observed
high-redshift galaxy UV luminosity functions \citep{Bouwens2015a,Bouwens2016,Oesch2018ApJ...855..105O},
the upper limit on the neutral hydrogen fraction at $z{\sim}5.9$ \citep{McGreer2015MNRAS.447..499M}, and
the CMB optical depth from {\it Planck} satellite \citep{Planck2016A&A...596A.108P}.
We calculate the 21-cm power spectra (PS) from this model, including
telescope noise corresponding to a 1000-hour integration with SKA1-low and moderate foreground avoidance.
These mock observations are then fed to the {\tocmmc} driver \citep{Greig2015MNRAS.449.4246G}, upgraded to allow for nested sampling \citep{Feroz_2008}, and used to constrain two models: (i) {\it 2pop}, including both MCGs and their massive atomic-cooling galaxy (ACG) counterparts; and (ii) {\it 1pop}, considering only ACGs.

We note that the ${\it 1pop}$ model is able to partially compensate for the missing population of MCGs by preferring a steeper stellar mass function (smaller $\alpha_\ast$) and a more X-ray luminous population of HMXBs (higher $L_{\rm X<2keV/\dot{\odot}}^{\rm atom}$).  However, without a transient population of MCGs, the more biased galaxies in the {\it 1pop} model result in a somewhat more rapid evolution of cosmic milestones, with a higher PS during the epoch of heating.

We quantify the preference of the mock observation for the more sophisticated galaxy model using the Bayesian evidence.
We obtain
$\ln\left[P(\mathscr{O}|\mathscr{M})\right]=-9.18\pm0.32$ and a maximum likelihood of $\ln\left[P(\mathscr{O}|\theta_{\rm max},\mathscr{M})\right]=-7.92$ for {\it 2pop}. These, compared to the {\it 1pop} result (i.e. $-33.01\pm0.25$ and $-20.44$), indicate a ${>}99.3\%$ probability of {\it 2pop} being preferred over {\it 1pop} by the data (i.e. the mock 21-cm PS) according to the Jeffreys' scale \citep{Jeffreys1939thpr.book.....J}. Thus if minihalo-hosted galaxies indeed have a significant impact on high-redshift IGM properties (as would be the case if the timing of the EDGES signal is proven to be cosmological \citealt{Bowman2018}), we should be able to indirectly infer their existence and their properties from upcoming 21-cm observations.

More generally, our study showcases how upcoming 21-cm measurements can be used to discriminate against uncertain galaxy formation models, of varying complexity (see also \citealt{Binnie_2019}).  Although we used two simplified, nested models here, the same analysis can be applied to even more sophisticated galaxy models (e.g. \citealt{Moster2013MNRAS.428.3121M,Sun2015,Mutch2016,Ma2018,Tacchella2018ApJ...868...92T,Behroozi2019MNRAS.488.3143B,Yung2019MNRAS.490.2855Y}).  The need for additional complexity can be directly tested via the Occam's razor factor of the Bayesian evidence, by adding additional model parameters until the evidence is maximized.

\section*{Acknowledgements}
This work was supported by the European Research Council (ERC) under the 
European Union’s 
Horizon 2020 research and innovation programme (AIDA -- \#638809). The results 
presented here
reflect the authors’ views; the ERC is not responsible for their use. Parts of 
this research 
were supported by the Australian Research Council Centre of Excellence for All 
Sky 
Astrophysics in 3 Dimensions (ASTRO 3D), through project \#CE170100013. This 
work was performed on the OzSTAR national facility at Swinburne University of 
Technology. OzSTAR is funded by Swinburne University of Technology and the 
National Collaborative Research Infrastructure Strategy (NCRIS). 
JP was supported in part by a KIAS individual Grant (PG078701) at Korea Institute for Advanced Study.

\section*{Data availability}
The data underlying this article will be shared on reasonable request to the corresponding author.

\bibliographystyle{\dir mn2e}
\bibliography{reference}


\bsp
\label{lastpage}
\end{document}